\documentstyle[aps,preprint,tighten]{revtex}
\begin{document}

\draft

\preprint{{\sl May 1995} \hskip 11.5cm SNUTP-95/054}

\title{Magnetic moments of heavy baryons in the Skyrme model}

\author{Yongseok Oh}

\address{Department of Physics, National Taiwan University,
	 Taipei, Taiwan 10764, Republic of China}

\author{Byung-Yoon Park\footnote{On leave of absence from
         Department of Physics, Chungnam National University,
         Daejeon 305-764, Korea}}

\address{Institute for Nuclear Theory, University of Washington,
	 Seattle, WA98195, U.S.A.}

\maketitle

\begin{abstract}
We calculate the magnetic moments of heavy baryons in the Skyrme model
in the limit of infinite heavy quark mass. We show that the Skyrme model
yields the same limit as the nonrelativistic quark model when heavy vector
mesons are treated properly. The essential role of the magnetic moment
coupling terms in the electromagnetic interactions of heavy mesons is
discussed.
\end{abstract}


The structure of hadrons containing a single heavy quark (or antiquark)
becomes independent of the spin and flavor of the heavy quark as its
mass ($m_Q^{}$) becomes sufficiently larger than the typical scale of
the strong interactions. \cite{IW89} A consequence of such heavy quark
symmetries can be found in the spectrum of heavy hadrons; that is, the
hadrons come in degenerate doublets \cite{IW91} with total spin
$j_\pm = j_\ell \pm \frac12$ (unless $j_\ell = 0$) with $j_\ell$ being
the total angular momentum of the light degrees of freedom. In the Skyrme
model (SM), heavy baryons with a single heavy flavor are described by
bound states of heavy mesons and a soliton of a chiral lagrangian. \cite{RRS}
Recently, it has been shown that such heavy quark symmetry can be
consistently incorporated into this picture by introducing heavy vector meson
fields. \cite{JMW,OPM3,JMW2,HBM2,HBM3,MOPR} There, the heavy vector mesons
play an essential role in making the  $\Sigma^*_Q$ and $\Sigma_Q^{}$ states
become degenerate in the infinite mass limit.

The magnetic moments also provide important informations on the baryon
structure. Although there are no experimental data  for the heavy baryon
magnetic moments, naive predictions have been made in the phenomenological
models such as nonrelativistic quark model (NRQM) \cite{NRQM,KP}, bag model
\cite{BM} and Skyrme Model \cite{OMRS,BDRS}.  In NRQM, the magnetic moment
of hadrons can be read off from their wave functions as a vector sum of the
contributions from their constituent quarks. The resulting magnetic moments
with arbitrary number of colors ($N_c = 2k+1$) are summarized in Table I for
the heavy baryons of our concern. There, $\mu_u^{}$, $\mu_d^{}$ and $\mu_Q^{}$
denote the corresponding magnetic moment of the constituent quarks. Assuming
that they are given by the charge-to-mass ratio, the magnetic moment of the
heavy quark $\mu_Q$ goes to 0 as $m_Q$ goes to infinity.

In the Skyrme model, the bound state approach has been shown to work well
in reproducing magnetic moments of the strange hyperons. \cite{OMRS,KMetc}
It is further applied to calculating the magnetic moments of charm
baryons \cite{OMRS} and bottom baryons \cite{BDRS}. Surprisingly, the Skyrme
model predictions are qualitatively very similar to those of NRQM. In SM,
the magnetic moments of heavy baryons are expressed in terms of four
{\it calculable\/} quantities, $\mu_s^{sol}$, $\mu_v^{sol}$, $\mu^{hm}_s$
and $\mu^{hm}_{v}$, as given in Table I. The first two quantities come from
the soliton configuration, while the rest two come from the bound heavy
mesons and {\it vanish\/} when the heavy quark (thus the heavy mesons)
becomes infinitely heavy. (See Ref. \cite{OMRS} for the details on these
quantities.) That is, in their works, both $\mu_s^{hm}$ and $\mu_v^{hm}$ are
of order of $1/m_Q$ while they are of order of 1 in $1/N_c$ counting. On the
other hand, comparing the NRQM and SM results, one can easily find the
relations between these quantities and the constituent quark magnetic moments
as
\begin{equation} \renewcommand{\arraystretch}{1.2} \begin{array}{cc}
\mu^{sol}_{s} = \mu_u^{} + \mu_d^{}, &
\mu^{hm}_{s} = 2 \mu_Q, \\
\mu^{sol}_{v} = \frac{2k+3}4 (\mu_u^{} - \mu^{}_d ), \qquad &
\mu^{hm}_{v} = - \frac14 (\mu_u^{} - \mu^{}_d ).
\end{array}
\label{rel}\end{equation}
{}From these relations, one can see that as far as the $1/N_c$ order counting
is concerned, $\mu_s^{hm}$ and $\mu_v^{hm}$ of Refs.\cite{OMRS,BDRS} are
consistent with NRQM.  But there seems to be a discrepancy in $1/m_Q$ order
counting between the two models; NRQM predicts $\mu_v^{hm}$ to be
${\cal O}(N_c^0 m_Q^0)$ while it is ${\cal O}(N_c^0 m_Q^{-1})$ in SM.
($\mu_s^{hm}$ is consistently of $1/m_Q$ order in both models.) However, in
Refs. \cite{OMRS,BDRS}, the calculations have been done in the model \cite{RRS}
where the heavy vector meson fields are integrated out in favor of the heavy
pseudoscalar meson field, which breaks the heavy quark symmetry seriously.
So it will be interesting to see whether the above mentioned discrepancy
is an artifact of such an approximation. In this paper, we show that the
SM and the NRQM have the {\it same\/} infinite heavy quark mass
limit of the heavy baryon magnetic moments when the heavy vector mesons are
treated properly.

We will work with the effective lagrangian constructed by Wise \cite{Wise}
for the heavy mesons interacting with Goldstone bosons, which reads
\begin{equation}
{\cal L} = {\cal L}_\Sigma^{(0)}
- i v^\mu \mbox{Tr} ({\cal D}^{(0)}_\mu H \bar{H})
- g \mbox{Tr}( H \gamma_5 \gamma^\mu {\cal A}^{(0)}_\mu \bar{H}).
\label{Lag0}\end{equation}
Here, ${\cal L}_\Sigma^{(0)}$ denotes the chiral lagrangian for the
Goldstone boson fields $\Sigma = \exp(i\bbox{\tau}\cdot\bbox{\pi}/f_\pi)$:
\begin{equation}
{\cal L}_\Sigma^{(0)} = \frac{f^2_\pi}{4} \mbox{Tr}
   (\partial_\mu \Sigma^\dagger \partial^\mu \Sigma) + \cdots,
\end{equation}
where $f_\pi$ is the pion decay constant and higher derivative terms are
abbreviated by the ellipses. The pseudoscalar ($j^\pi = 0^-$) and vector
($1^-$) heavy mesons fields, $P$ and $P^*_\mu$, are combined into a
$4\times 4$ matrix field $H(x)$,
\begin{equation}
H(x) = \frac{1 + v_\mu \gamma^\mu}{2}
( P_v^{} \gamma_5 - P^*_{v,\mu} \gamma^\mu),
\end{equation}
where the subscript $v$ of the field operator denotes that they are the
fields moving with a four-velocity $v_\mu^{}$. The field $H(x)$ has an
antidoublet structure in the isospin space, and transforms as
\begin{equation}
H \stackrel{\chi}{\longrightarrow} H h^\dagger,
\end{equation}
under the chiral transformation, while
\begin{equation}
\Sigma \stackrel{\chi}{\longrightarrow} L \Sigma R^\dagger,
\hskip 3mm
\xi\equiv \sqrt{\Sigma} \stackrel{\chi}{\longrightarrow}
	     L \xi h^\dagger = h \xi R^\dagger,
\end{equation}
with $L\in SU_L(2)$, $R\in SU_R(2)$ and $h$ being an $SU(2)$ matrix
depending on $L$, $R$ and $\xi$. The chiral covariant derivative
${\cal D}^{(0)}_\mu$ acting on $H$ is defined as
\begin{equation}
{\cal D}^{(0)}_\mu H = \partial_\mu H + H {\cal V}^{(0)\dagger}_\mu .
\end{equation}
The vector and axial vector fields, ${\cal V}^{(0)}_\mu$
and ${\cal A}^{(0)}_\mu$, are defined in terms of $\xi$ as
\begin{equation} \renewcommand{\arraystretch}{1.2} \begin{array}{l}
{\cal V}^{(0)}_\mu = \frac12 ( \xi^\dagger \partial_\mu \xi
	    + \xi \partial_\mu \xi^\dagger ), \\
{\cal A}^{(0)}_\mu = \frac{i}2 ( \xi^\dagger \partial_\mu \xi
	    - \xi \partial_\mu \xi^\dagger ).
\end{array} \end{equation}
The superscript `$(0)$' is adopted to distinguish the objects from those
after electromagnetic couplings. Finally, $g$ is a universal constant
for the heavy meson couplings to the pions.

Referring the details to Ref. \cite{OPM3}, we first briefly describe the
bound state approach. With a suitable stabilizing term, the nonlinear
lagrangian ${\cal L}_\Sigma^{(0)}$ supports a classical soliton solution in
the form of $ \Sigma_0^{}(\bbox{r}) = \exp [i\bbox{\tau}\cdot\hat{\bbox{r}}
F(r)]$ with $F(r)$ satisfying the boundary conditions $F(0)=\pi$ and
$F(r) \stackrel{r\rightarrow\infty}{\longrightarrow} 0$. It provides
static potentials to the heavy mesons so that they form a bound object
which carries a baryon number due to the soliton configuration and a
heavy flavor coming from the bound heavy mesons. In $m_Q^{}\rightarrow
\infty$ limit, the heavy mesons just sit at the center of the soliton and
the binding energy of this system can be naively estimated as
$\frac32 g F^\prime(0)$ with $F^\prime(0)$ being the slope of $F(r)$ at
the center.

Because of the hedgehog configuration the isospin ($\bbox{I}_h$) and the
angular momentum ($\bbox{L}$) of the heavy mesons become correlated, while
the heavy quark spin ($\bbox{S}_Q^{}$) decouples as a consequence of the
heavy quark symmetry. Thus, the soliton--heavy-meson bound states come out
as eigenstates of the `light quark grand spin' [$\bbox{K}_\ell^{}\equiv
(\bbox{S} - \bbox{S}_Q^{}) + \bbox{L} + \bbox{I}_h$ with the heavy meson
spin $\bbox{S}$], heavy quark spin and parity. For example, the wave
function of the eigenmode with $k_\ell\!=\!0$, $s_Q\!=\!\pm\frac12$
which is the lowest bound state is obtained as
\begin{equation}
H_{0,0,\pm\frac12}^{}(\bbox{r}) =  f(r) \frac{1+\gamma_0}{8\sqrt{2\pi}}
\phi_{\pm} [\gamma_5 - \bbox{\gamma}\cdot\bbox{\tau}]
(\bbox{\tau}\cdot\bbox{r}),
\end{equation}
where $f(r)$ is a radial function that is strongly peaked at the origin
and $\phi_{\pm}$ is the isospin basis for the antidoublet structure of
the heavy mesons. (See Refs. \cite{OPM3,MOPR} for details.) The heavy
meson field can be expanded in terms of these eigenmodes as
\begin{equation}
H(x) = \sum_{n} H_{n}(\bbox{r}) e^{-i\varepsilon_{n} t} a_{n},
\label{Hexp}\end{equation}
with the heavy meson annihilation operator $a_{n}^{}$
($n=k_\ell,\; k_{\ell,3}^{},\; s_Q^{} ).$
We will denote a single-particle Fock state as
$|n\rangle = a^\dagger_n |\mbox{vac}\rangle$, where the heavy mesons
occupy the corresponding bound state of the specified quantum numbers.

The quantization can be done by introducing collective coordinates
to the zero modes associated with the invariance under simultaneous
isospin rotation of the soliton field together with the heavy meson
field:
\begin{equation} \renewcommand{\arraystretch}{1.2} \begin{array}{c}
\xi(\bbox{r},t) = C(t) \xi_0(\bbox{r}) C^\dagger(t), \\
H(\bbox{r},t) = H_{\mbox{\scriptsize bf}}(\bbox{r},t) C^\dagger(t).
\end{array} \label{colrot}\end{equation}
Here, $H_{\mbox{\scriptsize bf}}$ represents the heavy meson field
in the rotating frame. Assuming sufficiently slow collective rotation,
we can expand it in terms of the unchanged classical eigenmodes
(\ref{Hexp}). In this collective coordinate quantization scheme, the
isospin of the heavy meson field is transmuted into the part of the
spin; the isospin operator $\bbox{I}$ and the spin operator $\bbox{J}$
of the soliton--heavy-meson bound system are given by
\begin{equation} \renewcommand{\arraystretch}{1.2} \begin{array}{l}
I_a^{} = D_{ab}^{}(C) R_b^{}, \\
\bbox{J} = \bbox{R} + \bbox{K}_{\mbox{\scriptsize bf}}.
\end{array}\end{equation}
Here, $\bbox{R}$ is the `rotor-spin' operator associated with the
collective rotation,
$D^{}_{ab}(C)[\equiv \frac12\mbox{Tr}(\tau_a^{} C \tau_b^{} C^\dagger)]$
is the $SU(2)$ adjoint representation associated with the collective
variables and $\bbox{K}_{\mbox{\scriptsize bf}}$ is the grand spin
operator of the heavy meson fields in the isospin co-moving system.
The wave functions of the rotor spin states are the Wigner $D$-functions,
$\sqrt{2i+1} D^{(i)}_{m_1m_2}(C)$ which satisfy
\begin{equation} \begin{array}{rcl}
\bbox{R}^2 D^{(i)}_{m_1,m_2}(C) &=& i(i+1) D^{(i)}_{m_1,m_2}(C), \\
{R}_3^{} D^{(i)}_{m_1,m_2}(C) &=& -m_2^{} D^{(i)}_{m_1,m_2}(C), \\
{I}_3^{} D^{(i)}_{m_1,m_2}(C) &=& m_1^{} D^{(i)}_{m_1,m_2}(C).
\end{array} \end{equation}

Since the heavy quark spin decouples, it is convenient to classify the
heavy baryons by the spin of the light degrees of freedom. The
corresponding operator $\bbox{J}_\ell^{}$ is given by
\begin{equation}
\bbox{J}_\ell^{} = \bbox{J} - \bbox{S}_Q^{} = \bbox{R} + \bbox{K}_\ell.
\end{equation}
The heavy baryon states with quantum numbers $j_\ell^{}$, $j_{\ell,3}^{}$
(spin of light degrees of freedom) and $i$, $i_3^{}$ (isospin) are
obtained by linear combinations of direct products of the rotor-spin
states $|i;m_1^{},m^{}_2\}$ and the single-particle Fock state
$|k^\pi_\ell, k_{\ell,3}^{}, s_Q^{}\rangle$. For example, $j_\ell^{}$=0
and $j_\ell^{}$=1 states are obtained by combining $i$=0 and 1 rotor-spin
states to $k_\ell^{}$=0 heavy meson bound state, respectively:
\begin{equation} \renewcommand{\arraystretch}{1.0} \begin{array}{l}
|j^{}_\ell\!=\!0, 0, s_Q^{} ; i\!=\!0, 0 \rangle\!\rangle
= D^{(0)}_{0,0} |k_\ell\!=\!0, 0; s_Q\rangle, \\
|j^{}_\ell\!=\!1,\: m_\ell^{}, s_Q^{} ; i\!=\!1, i_3^{} \rangle\!\rangle
= \sqrt3 D^{(1)}_{i_3,-m_\ell} |k_\ell\!=\!0, 0; s_Q\rangle.
\end{array} \label{hbs} \end{equation}
Conventional $\Lambda_Q^{}$, $\Sigma^{}_Q$ and $\Sigma^*_Q$ states are
obtained by combining the heavy quark spin and $j_\ell^{}$.

The magnetic moments of the heavy baryons can be obtained by taking
the expectation value of the corresponding operator
\begin{equation}
\bbox{\mu} =  {\textstyle\frac12} \int\!\! d^3r \:
 \bbox{r} \times \bbox{j}^{em} ,
\label{mmop}\end{equation}
with respect to the states given in Eq. (\ref{hbs}). Here, $\bbox{j}^{em}$
is the electromagnetic current, which can be derived by gauging the
electromagnetic $U(1)$ symmetry of the lagrangian. Under the electromagnetic
$U(1)$ transformation, the chiral field and the heavy meson field transform
as
\begin{equation} \renewcommand{\arraystretch}{1.4} \begin{array}{l}
\displaystyle
\Sigma \stackrel{U(1)}{\longrightarrow} e^{i {\cal Q}\lambda} \,
\Sigma \, e^{-i{\cal Q}\lambda}, \\
\displaystyle
H \stackrel{U(1)}{\longrightarrow} e^{i {\cal Q}^{\prime} \lambda} \,
H \, e^{-i{\cal Q}\lambda} ,
\end{array} \end{equation}
and similar equation for $\xi$. Here, ${\cal Q}$ is the charge matrix
associated with the light quark doublet, ${\cal Q}$ = diag($\frac23,
-\frac13)$,
and ${\cal Q}^{\prime}$ is the charge of the
heavy quark $Q$ [${\cal Q}^\prime(Q=c) =\frac23$, ${\cal Q}^\prime(Q=b)
= -\frac13$]. The minimal coupling of the electromagnetic field
$\alpha_\mu$ can be achieved by replacing all the plain derivatives in
the lagrangian (\ref{Lag0}) by the corresponding covariant derivatives:
\begin{equation}
\renewcommand{\arraystretch}{1.2} \begin{array}{l}
\partial_\mu \Sigma \Rightarrow
D_\mu \Sigma = \partial_\mu \Sigma - i e \alpha_\mu [{\cal Q}, \Sigma], \\
{\cal V}^{(0)}_\mu \!\Rightarrow
{\cal V}_\mu \!=\! \frac12( \xi^\dagger D_\mu \xi + \xi D_\mu \xi^\dagger)
 \!=\! {\cal V}^{(0)}_\mu \!\!+\!
  i e \alpha_\mu ({\cal Q}\! -\! {\cal Q}_V^{} ), \\
{\cal A}^{(0)}_\mu \!\Rightarrow
{\cal A}_\mu \!=\! \frac{i}{2} (\xi^\dagger D_\mu \xi\!
 -\! \xi D_\mu \xi^\dagger)
 \!=\! {\cal A}^{(0)}_\mu \!\! + \! e \alpha_\mu {\cal Q}_A^{}, \\
{\cal D}^{(0)}_\mu H \Rightarrow
{\cal D}_\mu H = \partial_\mu H \!+\! H {\cal V}_\mu^\dagger
 \!-\! ie \alpha_\mu ( {\cal Q}^\prime H \!-\! H {\cal Q}) \\
\hskip 6.8em = 
{\cal D}^{(0)}_\mu H
\!-\! i e \alpha_\mu ( {\cal Q}^\prime H \!-\! H {\cal Q}_V^{} ), \\
\end{array}
\end{equation}
where
${\cal Q}_V = \frac12(\xi^\dagger {\cal Q} \xi + \xi {\cal Q} \xi^\dagger)$
and
${\cal Q}_A = \frac12(\xi^\dagger {\cal Q} \xi - \xi {\cal Q} \xi^\dagger)$.
Note that the $SU(2)$ flavor symmetry is broken by the electromagnetic
interactions. However, the charge operator ${\cal Q}$ has an equal
mixture of $\bf{3}_L$ and $\bf{3}_R$ since the electromagnetic interactions
conserve parity.

However, such a minimal coupling cannot incorporate the radiative transition
like $P^* \rightarrow P\gamma$, because ${\cal V}_\mu$ $({\cal A}_\mu)$
contains only an even (odd) number of pions interacting electromagnetically;
that is, the kinetic term of Eq. (\ref{Lag0}) gives rise to contact terms
with one photon and even-number pion emissions, while the interaction term
yields those with odd-number of pions. The lowest order interaction term
that contributes to $P^* \rightarrow P\gamma$ is \cite{CCLLYY,CG,ABJLMRSW}
\begin{equation}
{\cal L}_{\text{mag}} = \frac{\kappa}{2} F_{\mu\nu} \mbox{Tr}
  ( H \sigma^{\mu\nu} {\cal Q}_V^{} \bar{H}),
\label{Lem2} \end{equation}
where
$F_{\mu\nu}=\partial_\mu^{}\alpha_\nu^{} - \partial_\nu^{}\alpha_\mu^{}$
and $\sigma_{\mu\nu}^{} = \frac{i}{2} [\gamma_\mu^{}, \gamma_\nu^{}]$.
It comes from the ``anomalous" magnetic moment of the heavy vector
mesons due to their internal structure.\footnote{
   In Ref. \cite{CCLLYY}, the authors work with the conventional
   pseudoscalar and vector fields to describe the heavy mesons with
   finite masses. One can easily obtain Eq. (\ref{Lem2}) by substituting
   $$\begin{array}{l}
   P = e^{-i m_P^{} v\cdot x} \frac{1}{\sqrt{m_P^{}}} P_v^{}, \\
   P^*_\mu = e^{-i m_{P^*} v\cdot x} \frac{1}{\sqrt{m^{}_{P^*}}}
   P^*_{v,\mu},
   \end{array} $$
   into their lagrangian ${\cal L}^{(2)}_{PP^*}$ and keeping the
   leading order terms in $1/m_P$.
   In Ref. \cite{CG}, the ``bare" charge matrix ${\cal Q}$ is adopted
   instead of the ``dressed" one ${\cal Q}_V^{}$. Although both yield
   the same transition rates for the process $P^* \rightarrow P \gamma$
   at tree level, using ${\cal Q}_V$ seems to give the correct result
   in our work. There also can be a term $F_{\mu\nu} \mbox{Tr} (\bar H
   \sigma^{\mu\nu} {\cal Q}^\prime H)$ in ${\cal L}_{\text{mag}}$, but
   it is suppressed by $1/m_Q$. \cite{CCLLYY,CG}
} 
If the heavy mesons are so strongly bound that they have zero
bound-state radius, the magnetic moment of the heavy vector mesons will
take the canonical value $e/m_{P^*}^{}$ \cite{BH} and would vanish at the
infinite heavy meson mass limit. On the other hand, the NRQM provides a
simple prediction for $\kappa$ \cite{CCLLYY}:
\begin{equation}
\kappa = \frac{e}{2m_u^{}} = \textstyle\frac32\mu_u,
\label{kappa}\end{equation}
which is in the range of the fitted value given in Ref. \cite{CG}.

Now, the electromagnetic current can be directly read off from the lagrangian
(\ref{Lag0}) and (\ref{Lem2}) as
\begin{equation} \renewcommand{\arraystretch}{1.5}\begin{array}{rl}
j^\mu_{em} = \!\! & \displaystyle
\frac{1}{48\pi^2} \varepsilon^{\mu\nu\lambda\rho}
    \mbox{Tr} (\Sigma^\dagger \partial_\nu \Sigma \Sigma^\dagger
      \partial_\lambda \Sigma \Sigma^\dagger \partial_\rho \Sigma )
   + ie\frac{f^2_\pi}{2}
      \mbox{Tr} (\Sigma^\dagger {\cal Q} \partial^\mu \Sigma
           + \Sigma {\cal Q} \partial^\mu \Sigma^\dagger) + \cdots \\
  & - e{\cal Q}^\prime v^\mu \mbox{Tr}(H \bar{H})
   + e v^\mu \mbox{Tr} (H {\cal Q}_V \bar{H} )
   - eg \mbox{Tr} (H \gamma_5 \gamma^\mu {\cal Q}_A \bar{H} )
   + \kappa \partial_\nu [ \mbox{Tr} (H \sigma^{\mu\nu} {\cal Q}_V \bar{H})],
\end{array} \end{equation}
where we have included the ``baryon number" current as the isoscalar
component of the electromagnetic current coming from the soliton.
The ellipsis denotes the contributions of higher derivative terms in the
chiral lagrangian ${\cal L}^{(0)}_\Sigma$. The magnetic moment operator
can be obtained by substituting the space component of the
electromagnetic current into Eq. (\ref{mmop}) with the `rotating fields'
of Eq. (\ref{colrot}). Since $\bbox{v}$ is of order of $1/m_Q$ and the
bound heavy mesons are strongly peaked at the origin where
${\cal Q}_A$ vanishes, {\it the contribution of heavy mesons to the
magnetic moments comes only from the last term\/}. Finally, we are led to
\begin{equation} \renewcommand{\arraystretch}{1.5} \begin{array}{rl}
\mu_3^{} \!=\!\! & \mu^{sol}_s {R}_3^{} - 2 \mu_v^{sol} D_{33}(C) \\
  & -2\kappa \displaystyle \int\! d^3r \: \mbox{Tr}
    [ H_{\mbox{\scriptsize bf}} (\textstyle - \frac12 \sigma_3^{})
    C^\dagger {\cal Q}_V^{} C \bar{H}_{\mbox{\scriptsize bf}} ].
\end{array}\end{equation}
(See, {\it e.g.\/}, Ref. \cite{OMRS} for explicit forms of
$\mu^{sol}_{s,v}$.) When the expectation value of the last term with
heavy meson field operators is taken with respect to the single-particle
Fock state $|k_\ell\!=0,0,s_Q\rangle$ of Eq. (\ref{hbs}), the magnetic
moment operator is reduced to a form of
\begin{equation} \label{newmm}
\mu_3^{} \!=\! \mu^{sol}_s {R}_3^{} - 2 (\mu_v^{sol}
 + \mu_v^{hm}) D_{33}(C),
\end{equation}
where\footnote{
If one has used ${\cal Q}$ in Eq. (\ref{Lem2}) instead of ${\cal Q}_V$,
he would have obtained $\mu^{hm}_v = + \frac1{12} \kappa$.}
\begin{equation}
\mu_v^{hm} = - \textstyle \frac14 \kappa,
\label{muvm}\end{equation}
which acts on the rotor-spin states in Eq. (\ref{hbs}). In this formula, it
can be seen that the heavy mesons do not contribute to the isoscalar part of
the magnetic moment so that $\mu_s^{hm}$ is ${\cal O}(N_c^0 m_Q^{-1})$, which
is consistent with the NRQM. The expectation value of $R_3$ and $D_{33}(C)$
with respect to the rotor spin states $D^{(1)}_{m_1m_2}(C)$ are $-m_2^{}$ and
$m_1^{} m_2^{}/2$, respectively. The magnetic moments of the conventional
heavy baryons such as $\Sigma_Q^{}$ and $\Sigma^*_Q$ can be obtained by
multiplying a factor that appears in combining the heavy quark spin to
$j_\ell$,
which yields exactly the same results of Table I but with nonvanishing
$\mu^{hm}_v$ in the infinite heavy quark mass limit. Therefore,
${\cal L}_{\text{mag}}$ of Eq. (\ref{Lem2}) gives a contribution of
${\cal O}(N_c^0 m_Q^0)$ that was missing in the earlier calculations
\cite{OMRS,BDRS}. Note also that, with the NRQM estimation of $\kappa$ given
by Eq. (\ref{kappa}), we can reproduce the last relation of Eq. (\ref{rel}).
When the two parameters $\mu^{sol}_{s}$ and $\mu^{sol}_v$ are adjusted to fit
the nucleon magnetic moments, {\it both models\/} predict on the heavy baryon
magnetic moments as
\begin{equation} \begin{array}{ll}
\mu (\Lambda_Q) = 0, & \mu (\Sigma_Q^0) = \frac29 \mu_p, \\
\mu (\Sigma_Q^{+1}) = \frac89 \mu_p, \qquad
& \mu (\Sigma_Q^{-1}) = -\frac49 \mu_p,
\end{array}
\end{equation}
in the infinite heavy quark mass limit. [$\mu(\Sigma_Q^*)$'s are given by the
relation $\mu(\Sigma_Q^*) = \frac32 \mu(\Sigma_Q)$.]

As a summary we have shown that the Skyrme model could yield the same heavy
baryon magnetic moments as NRQM in the limit of the heavy quark mass going to
infinity if heavy vector mesons are treated  properly. This study tells us
that the ``anomalous" coupling term(s) in the Lagrangian for the
electromagnetic
transition of the heavy vector mesons can play a nontrivial role in the Skyrme
model calculations on these physical quantities. Their contributions to the
isovector part of the magnetic moments are  of the next leading order in
$1/N_c$ counting but of the leading order in the $1/m_Q$ expansion. Actually,
in the literature \cite{KMetc,mmSM}, the Skyrme model estimations on the baryon
magnetic moments have suffered from too small isovector part. The inclusion
of the vector mesons \cite{mmN} into the model, which has been expected to
cure the problem, could not solve this problem completely although it
improved the model predictions. So it will be interesting to see how much
the incorporation of the ``anomalous" coupling terms ({\it e.g.\/},
$K^*K\gamma$ and $\rho\pi\gamma$ terms) into the magnetic moment calculations
improves the results.

In this work, we have worked with finite $N_c$ but with infinite heavy quark
mass. In Refs. \cite{OP-M,Syr}, it was shown that the kinetic effects of the
heavy mesons give nontrivial corrections in the mass spectrum. In order to
have more realistic predictions on the magnetic moments of the heavy baryons
with finite masses, one should take into account the finite mass corrections
on these physical quantities. Such a work is in progress and the results will
be presented in a further publication \cite{OPM4}.

\vskip 1.0cm
We are very grateful to D.-P. Min, T.-S. Park and N.N. Scoccola for helpful
discussions and to M. Rho for the large $N_c$ interpretation of our result.
This work was supported in part by the National Science Council of ROC under
Grant No. NSC84-2811-M002-036 and in part by the Korea Science and Engineering
Foundation through the SRC program.


\begin{table}
\begin{center}
\caption{Magnetic moment of heavy baryons. The superscripts
($0$,$\pm 1$) are the isospin projections, therefore, $\Sigma_Q^{+1}$
means $\Sigma_b^+$ for $Q$=$b$ and $\Sigma_c^{++}$ for $Q$=$c$.}
\begin{tabular}{ccc}
Particle & NRQM ($N_c$=$2k+1$) \cite{KP} & SM \cite{OMRS} \\
\hline
$p$ & $\frac13 [ (k+3)\mu_u - k \mu_d ]$
  & $\frac12 \mu^{sol}_{s} + \frac23 \mu^{sol}_{v}$ \\
$n$ & $\frac13 [ (k+3)\mu_d - k \mu_u ]$
  & $\frac12 \mu^{sol}_{s} - \frac23 \mu^{sol}_{v}$ \\
$\Lambda_Q$ & $\mu_Q$ & $\frac12 \mu^{hm}_{s}$ \\
$\Sigma_Q^{+1}$ & $\frac13 [ (k+3)\mu_u - (k-1)\mu_d ] - \frac13 \mu_Q^{}$
  & $ \frac16 ( 4 \mu^{sol}_{s} - \mu^{hm}_{s} )
     + \frac23 ( \mu^{sol}_{v} + \mu^{hm}_{v} ) $ \\
$\Sigma_Q^{0}$ & $\frac23 (\mu^{}_{u} + \mu_d) - \frac13 \mu_Q^{}$
  & $ \frac16 ( 4 \mu^{sol}_{s} - \mu^{hm}_{s} ) $ \\
$\Sigma_Q^{-1}$ & $\frac13 [ (k+3)\mu_d - (k-1)\mu_u ] - \frac13 \mu_Q^{}$
  & $ \frac16 ( 4 \mu^{sol}_{s} - \mu^{hm}_{s} )
     - \frac23 ( \mu^{sol}_{v} + \mu^{hm}_{v} ) $  \\
$\Sigma_Q^{*+1}$ & $\frac12 [(k+3) \mu^{}_{u} - (k-1)\mu_d ] + \mu_Q^{}$
  & $ \mu^{sol}_{s} + \frac12 \mu^{hm}_{s}
     + ( \mu^{sol}_{v} + \mu^{hm}_{v} ) $ \\
$\Sigma_Q^{*0}$ & $ \mu^{}_{u} + \mu_d + \mu_Q^{}$
  & $ \mu^{sol}_{s} + \frac12 \mu^{hm}_{s} $  \\
$\Sigma_Q^{*-1}$ & $ \frac12 [ (k+3)\mu_d - (k-1)\mu_u ] + \mu_Q^{}$
  & $  \mu^{sol}_{s} + \frac12 \mu^{hm}_{s}
     - ( \mu^{sol}_{v} + \mu^{hm}_{v} ) $ \\
\end{tabular}
\end{center}
\end{table}


\begin{references}

\bibitem{IW89}
   M.B. Voloshin and M.A. Shifman,
      Sov. J. Nucl. Phys. 45 (1987) 292;
      Sov. J. Nucl. Phys. 47 (1988) 511; \\
   N. Isgur and M.B. Wise,
      Phys. Lett. B 232 (1989) 113;
      Phys. Lett. B 237 (1990) 527.

\bibitem{IW91}
   N. Isgur and M.B. Wise,
      Phys. Rev. Lett. 66 (1991) 1130.

\bibitem{RRS}
   C.G. Callan and I. Klebanov,
      Nucl. Phys. B 262 (1985) 365; \\
   M. Rho, D.O. Riska and N.N. Scoccola,
      Phys. Lett. B 251 (1990) 597;
      Z. Phys. A 341 (1992) 343.

\bibitem{JMW}
   E. Jenkins, A.V. Manohar and M.B. Wise,
      Nucl. Phys. B 396 (1993) 27.

\bibitem{OPM3}
   Y. Oh, B.-Y. Park and D.-P. Min,
      Phys. Rev. D 50 (1994) 3350.

\bibitem{JMW2}
   E. Jenkins and A.V. Manohar,
      Phys. Lett. B 294 (1992) 273; \\
   Z. Guralnik, M. Luke and A.V. Manohar,
      Nucl. Phys. B 390 (1993) 474.

\bibitem{HBM2}
   M.A. Nowak, M. Rho and I. Zahed,
      Phys. Lett. B 303 (1993) 130; \\
   H.K. Lee and M. Rho,
      Phys. Rev. D 48 (1993) 2329; \\
   H.K. Lee, M.A. Nowak, M. Rho and I. Zahed,
      Ann. Phys. (N.Y.) 227 (1993) 175.

\bibitem{HBM3}
   K.S. Gupta, M.A. Momen, J. Schechter and A. Subbaraman,
      Phys. Rev. D 47 (1993) 4835; \\
   A. Momen, J. Schechter and A. Subbaraman,
      Phys. Rev. D 49 (1994) 5970.

\bibitem{MOPR}
   For a review, see, for example,
   D.-P. Min, Y. Oh, B.-Y. Park and M. Rho,
      Int. J. Mod. Phys. E 4 (1995) 47.

\bibitem{NRQM}
   D.B. Lichtenberg,
      Phys. Rev. D 15 (1977) 345.

\bibitem{KP}
   G. Karl and J.E. Paton,
      Phys. Rev. D 30 (1984) 238.

\bibitem{BM}
   S.K. Bose and L.P. Singh,
      Phys. Rev. D 22 (1980) 773.

\bibitem{OMRS}
   Y. Oh, D.-P. Min, M. Rho and N.N. Scoccola,
	  Nucl. Phys. A 534 (1991) 493.

\bibitem{BDRS}
   M. Bj\"{o}rnberg, K. Dannbom, D.O. Riska and N.N. Scoccola,
	  Nucl. Phys. A 539 (1992) 662.

\bibitem{KMetc}
   J. Kunz and P.J. Mulders,
      Phys. Lett. B 231 (1989) 335;
      Phys. Rev. D 41 (1990) 1578; \\
   E.M. Nyman and D.O. Riska,
      Nucl. Phys. B 325 (1989) 593; \\
   D.-P. Min, Y.S. Koh, Y. Oh and H.K. Lee,
      Nucl. Phys. A 530 (1991) 698.

\bibitem{Wise}
   M.B. Wise,
      Phys. Rev. D 45 (1992) 2188.

\bibitem{CCLLYY}
   H.-Y. Cheng {\it et al.\/},
      Phys. Rev. D 47 (1993) 1030.

\bibitem{CG}
   P. Cho and H. Georgi,
      Phys. Lett. B 296 (1992) 408; (E) 300 (1993) 410.

\bibitem{ABJLMRSW}
   J.F. Amundson {\it et al.\/},
      Phys. Lett. B 296 (1992) 415.

\bibitem{BH}
   S.J. Brodsky and J.R. Hiller,
      Phys. Rev. D 46 (1992) 2141.

\bibitem{mmSM}
   G.S. Adkins, C.R. Nappi and E. Witten,
      Nucl. Phys. B 228 (1983) 552; \\
   G.S. Adkins and C.R. Nappi,
      Phys. Lett. B 137 (1984) 251.

\bibitem{mmN}
   U.-G. Meissner,
      Phys. Rep. 161 (1988) 213 and references therein.

\bibitem{OP-M}
   Y. Oh, B.-Y. Park and D.-P. Min,
      Phys. Rev. D 49 (1994) 4649; \\
   Y. Oh and B.-Y. Park,
      SNU preprint SNUTP-94/131, hep-ph/9501356,
      Phys. Rev. D 51 (in press).

\bibitem{Syr}
   J. Schechter and A. Subbaraman,
      Phys. Rev. D 51 (1995) 2311; \\
   J. Schechter, A. Subbaraman, S. Vaidya and H. Weigel,
      Syracuse preprint SU-4240-606, hep-ph/9503307 (1995).

\bibitem{OPM4}
   Y. Oh, B.-Y. Park and D.-P. Min, work in progress.

\end{references}
\end{document}